\documentclass[useAMS,usenatbib]{mn2e}
\usepackage{amssymb,amsmath,epsfig,times,natbib,longtable,color,graphicx}
\usepackage[draft]{hyperref}

\long\def\symbolfootnote[#1]#2{\begingroup%
\def\thefootnote{\fnsymbol{footnote}}\footnote[#1]{#2}\endgroup} 
\def\apj{ApJ}
\def\apjl{ApJ}
\def\aap{A\&A}
\def\aapr{A\&A~Rev.}
\def\mnras{MNRAS}
\def\pasp{PASP}
\def\pasj{PASJ}
\def\ssr{Space~Sci.~Rev.}
\def\nat{Nature}

\newcommand{\suzaku}{{\it Suzaku}}

\newcommand{\xmm}{{\it XMM-Newton}}

\makeatletter
 \def\hlinewd#1{%
   \noalign{\ifnum0=`}\fi\hrule \@height #1 \futurelet
    \reserved@a\@xhline}
\makeatother

\title[X-ray reflection in Ton S180]{X-ray reflection from the inner disk of the AGN Ton S180}

\author[M. L. Parker et al.]{M. L. Parker$^{1,2}$,\thanks{Email: mlparker@ast.cam.ac.uk}
J. M. Miller$^{3}$
and A. C. Fabian$^{2}$\\
 $^{1}$European Space Agency (ESA), European Space Astronomy Centre (ESAC), E-28691 Villanueva de la Ca\~{n}ada, Madrid, Spain\\
  $^2$Institute of Astronomy, Madingley Road, Cambridge, CB3 0HA, UK\\
  $^3$Department of Astronomy, University of Michigan, 1085 South University Avenue, Ann Arbor, MI 48109-1104, USA\\
}
\date{}

\begin{document}

\maketitle

\begin{abstract}
We analyse a long archival \xmm\ observation of the narrow-line Seyfert 1 galaxy Ton~S180, using the latest reflection models to explore the high quality X-ray spectrum. We find that the iron line is relatively narrow and sharp, and the soft excess is extremely smooth. We cannot find an acceptable reflection model that describes both components, and conclude that the soft excess cannot be produced by relativistic reflection. Fitting the 3--10~keV band with relativistic reflection to model the iron line strongly prefers low spin values ($<0.4$), with the exact value depending on the model and not well constrained. We then model the broad-band spectrum with a two-component Comptonization continuum plus relativistic reflection. This gives a much better fit than a pure reflection model, which again prefers a low spin value. The photon index of the reflection component is intermediate between the two Comptonization components, suggesting that both illuminate the disk a similar amount and therefore both contribute to the reflection. 
\end{abstract}


\section{Introduction}

Astrophysical black holes are famously described by just two parameters: their mass, $M$, and spin, $a$, with a third parameter, charge, tending rapidly to zero in any realistic environment. To understand the formation and growth of black holes, we therefore only need to measure these two parameters to build a complete picture of each object, and then measure enough objects to have accurate population statistics.
The leading method of estimating spin in active galactic nuclei (AGN) is the iron line or relativistic reflection method \citep{Fabian89}. This technique relies on establishing the relativistic distortion of narrow emission lines, produced when the inner accretion disk is illuminated by X-rays produced in the X-ray corona. Information about the spin, inclination, and geometry of the inner disk is imprinted into the line profile, and can be estimated by spectral fitting.

A common criticism of reflection fitting is the limited range of solutions found -- the observed distribution is heavily skewed towards maximal spin \citep{Reynolds14}. In the most up-to-date list of reflection spins \citep{Vasudevan16}, 17 of the 25 are lower limits, and two more have estimates with $0.9<a<0.998$. There are potentially multiple reasons for this: high spin AGN radiate more efficiently, and are thus brighter; they are more rapidly variable, and are therefore more likely to be observed; it is easier to constrain spin from a strongly relativistically broadened line profile than a narrower line, and less degenerate with cold, distant reflection; emission from within the innermost stable circular orbit (ISCO) may bias the fitting towards high spins \citep{Reynolds08}; fitting the ambiguous `soft excess' with reflection when it is not appropriate may give incorrect results; relativistic reflection may not be the correct model to fit a given spectrum; and finally the true distribution of AGN may be skewed towards high spin \citep[e.g.][]{Dotti13}. 
Measuring low spin values in AGN is thus a crucial test of both the reflection method itself, and of the distribution of black hole spins.

A key issue with measuring spin using reflection is the nature of the soft excess. It is possible that the soft excess is due to the blurred low energy emission lines produced by reflection \citep{Crummy06}, in which case it should be possible to constrain the spin from fitting the soft excess. However, if the soft excess is instead due to another process, such as Comptonization of disk photons, fitting it with reflection will strongly bias the spin result \citep[see discussion in][]{Reynolds14}.
For example, \citet{Lohfink12} find two solutions to the spin of Fairall~9, one with near maximal spin when the soft excess is modelled with a second relativistic reflection component and one with more moderate spin when the soft excess is modelled with Comptonization. In the ideal scenario, the spin can be measured from the iron line alone \citep[e.g.][]{Parker17_iras13224}, and should be consistent with that found using the full broad-band spectrum \citep{Fabian13_iras}.

Ton~S180 is one of the most variable AGN in the \xmm\ archive \citep{Ponti12}, and has previously been found to have high spin from reflection modelling \citep{Nardini12, Walton13_suzaku}. However, the soft excess in Ton~S180 is unusual. \citet{Vaughan02} describe it as `a smooth, featureless excess', and there is no clear break indicating the end of the soft excess and start of the power-law continuum. \citeauthor{Vaughan02} successfully fit the broad-band spectrum with a double power-law model, finding a poor fit with a power-law plus black-body model. Similarly, \citet{Murashima05} find a best fit double power-law model to archival \emph{ASCA}, \emph{RXTE} and \xmm\ data. \citet{Crummy06} also note the unusual shape of the soft excess, which extends to higher energies than any other source in their sample and cannot be well fit with their reflection model. 

Interestingly \citet{Takahashi10}, using \suzaku\ data, find that an additional component is required as well as relativistic reflection to model the soft excess, and then find a significant degree of truncation ($R_\mathrm{in}\sim25R_\mathrm{G}$). However, their fits all find an edge-on inclination (90 degrees, i.e. through the disk and torus), so it is not clear how reliable this conclusion is.

In this paper we present the case for a low spin interpretation of the X-ray spectrum of the AGN Ton~S180, using archival \xmm\ data to constrain the spin using the iron line and compare it with the results from previous broad-band fitting. 

\section{Observations and Data Reduction}
\label{section_datareduction}

We reduce the \xmm\ data using the latest Science Analysis Software (SAS, version 15.0.0). We use the longest publicly available archival observation (obs ID 0764170101, from 3rd July 2015). For the EPIC-pn \citep{Struder01}, we extract event files using the \textsc{epproc} ftool. We extract only single and double events, and filter for background flares. We extract source photons from 40$^{\prime\prime}$ circular regions, and background photons from larger circular regions on the same chip, avoiding the high copper background region and contaminating sources. We bin the spectra to oversample the data by a factor of 3, and to a signal-to-noise ratio of 6, after background subtraction, to ensure the viability of $\chi^2$ statistics. All fits to the EPIC-pn data use \textsc{xspec} version 12.9.1a \citep{Arnaud96}. We use the RGS data from the same observation, fitting with the standard spectra, background and response files from the pipeline processing. All RGS fits use \textsc{spex} version 3.02.00 \citep{Kaastra96}.

\section{Results}

\subsection{RGS}
We use the RGS data to identify any potential narrow emission or absorption features in the data. We initially fit two \emph{hot} components, to model the interstellar absorption (both Galactic and source), one at zero shift with respect to the Milky Way, and one at zero shift with respect to Ton S180 ($z=0.062$). These model ISM absorption that might otherwise be confused for a warm absorber. Abundances were frozen at solar values in each \emph{hot} component, and the continuum was assumed to be a simple power-law. This fit achieves $\chi^2 = 2090.8$ for 1867 degrees of freedom, with no strong evidence of  emission or absorption lines. High and low bins appear to be random and evenly distributed, and seldom correspond when the RGS1 and RGS2 residuals are compared. The ISM absorber at rest in the Ton S180 frame has a 90\% confidence upper limit of $4.8\times10^{19}$~cm$^{-2}$, and the Galactic absorber has an upper limit of $3.9\times10^{19}$~cm$^{-2}$

Next, we add a \emph{pion} component to the fit, to search for any evidence of warm absorption. \emph{Pion} is a photoionized absorption code, which includes a number of intermediate charge states that are absent in many other models, and also has the advantage of self-consistently adjusting the input luminosity and spectrum to match the local continuum model.
A number of warm absorbers have a turbulent velocity of 300~km~s$^{-1}$ \citep[e.g.][]{lee13}, so this value was fixed in fits with \emph{pion}. As with the other components, solar abundances were assumed. Fits with \emph{pion} do not register significant improvements to the fit (e.g. $\Delta\chi^2$ of 10, for 3 degrees of freedom). We scan in velocity space to search for any evidence of warm absorption, using a fiducial ionization of $\log(\xi)=2$ erg~cm~s$^{-1}$ and searching between 0 and 5000~km~s$^{-1}$. No significant absorption is detected, and we find an upper limit on the column density of 1.2$\times10^{20}$~cm$^{-2}$.

Finally, we perform a blind line scan over the whole RGS band in \textsc{xspec} with a narrow Gaussian line ($\sigma=1$~eV), using a power-law continuum and Galactic absorption modelled with \emph{TBnew} \citep{Wilms00}. We allow the line to have both positive and negative normalizations. The highest significance we find is a negative residual at $\Delta\chi^2$ of 13, at a rest frame wavelength of 25.2~\AA. When the extremely large number of trials (due to the very high resolution of the RGS) are taken into account, the significance of this is negligible (we would expect to find $\sim0.6$ such features in a random spectrum). We conclude that there is no evidence for any absorption in the RGS spectrum.

\subsection{Iron band fitting}
\label{subsec_ironband}

\begin{figure}
\centering
\includegraphics[width=\linewidth]{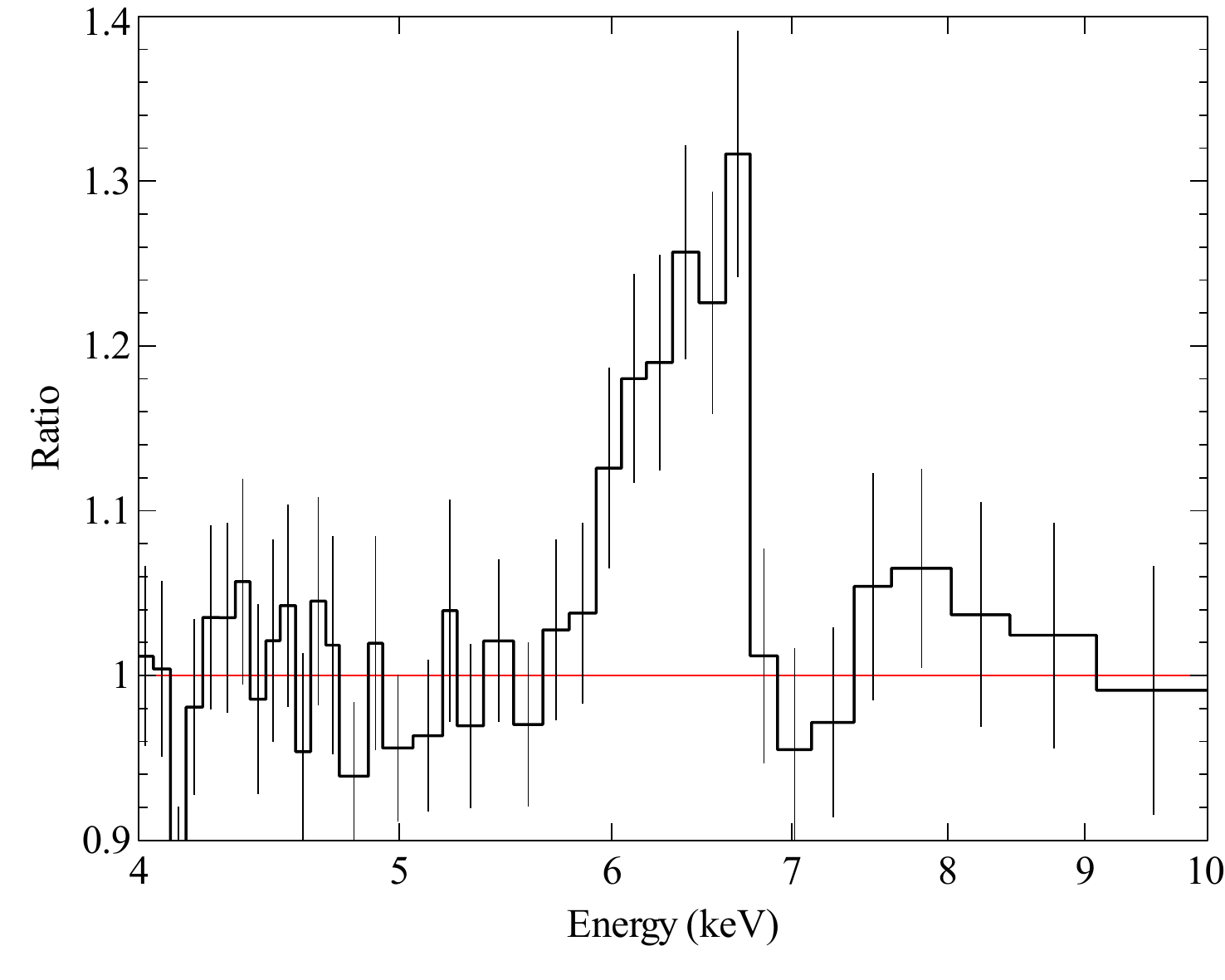}
\caption{Zoom of the iron line profile. The line is shown in the residuals to a power-law, fit from 3-5 and 7-10~keV. The line is clearly skewed, but relatively narrow, suggesting a low spin. Data are rebinned in \textsc{xspec} for clarity. Energies are in the observer's frame.}
\label{fig_lineprofile}
\end{figure}

We initially fit the EPIC-pn spectrum from 3--10~keV, so that the soft excess is excluded from the fit. This ensures that the spin constraint is being driven by the iron line, at the expense of making estimates of the reflection fraction and iron abundance unreliable \citep{Parker15_cygx1}. The iron line profile (shown in Fig.~\ref{fig_lineprofile}) is relatively narrow compared to many of the other NLS1 sources, and the blue side of the line is quite sharp. It is possible that a profile like this could also be caused by partial-covering absorption \citep[see e.g. review by][]{Turner09Rev}. However, to produce just this feature would require a high covering fraction and a very high column density, which we think is unlikely in a source with no other evidence for absorption. We therefore prefer to consider only the relativistic reflection interpretation in this work, and do not explore alternative models. We discuss the partial-covering interpretation further in \S~\ref{section_discussion}.

As in \citet{Parker14_mrk335}, we fit three models to the data: \emph{relxill} \citep{Garcia13}, \emph{relxilllp} (the lamp-post geometry version of \emph{relxill} -- \emph{reflionx} and \emph{relxill} do not assume any particular geometry), and \emph{reflionx} \citep{Ross05}, convolved with \emph{relconv} \citep{Dauser10}. By using these three models, we can be sure that our conclusions are robust to choice of atomic physics or assumed geometry. \emph{Reflionx} and \emph{relxill} assume slightly different atomic physics, and \emph{reflionx} is angle-averaged, meaning that there can be systematic differences in measured inclination \citep{Middleton16}. We use a power-law continuum (this is built in to the \emph{relxill} models), and Galactic absorption of $1\times10^{20}$~cm$^-2$, modelled with \emph{tbnew}. We use \textsc{vern} cross sections \citep{Verner96} and \textsc{wilm} abundances \citep{Wilms00}. The main aim of fitting these models is to establish the spin parameter and inclination of the inner accretion disk ($a$ and $i$, respectively).

\begin{figure}
\centering
\includegraphics[width=\linewidth]{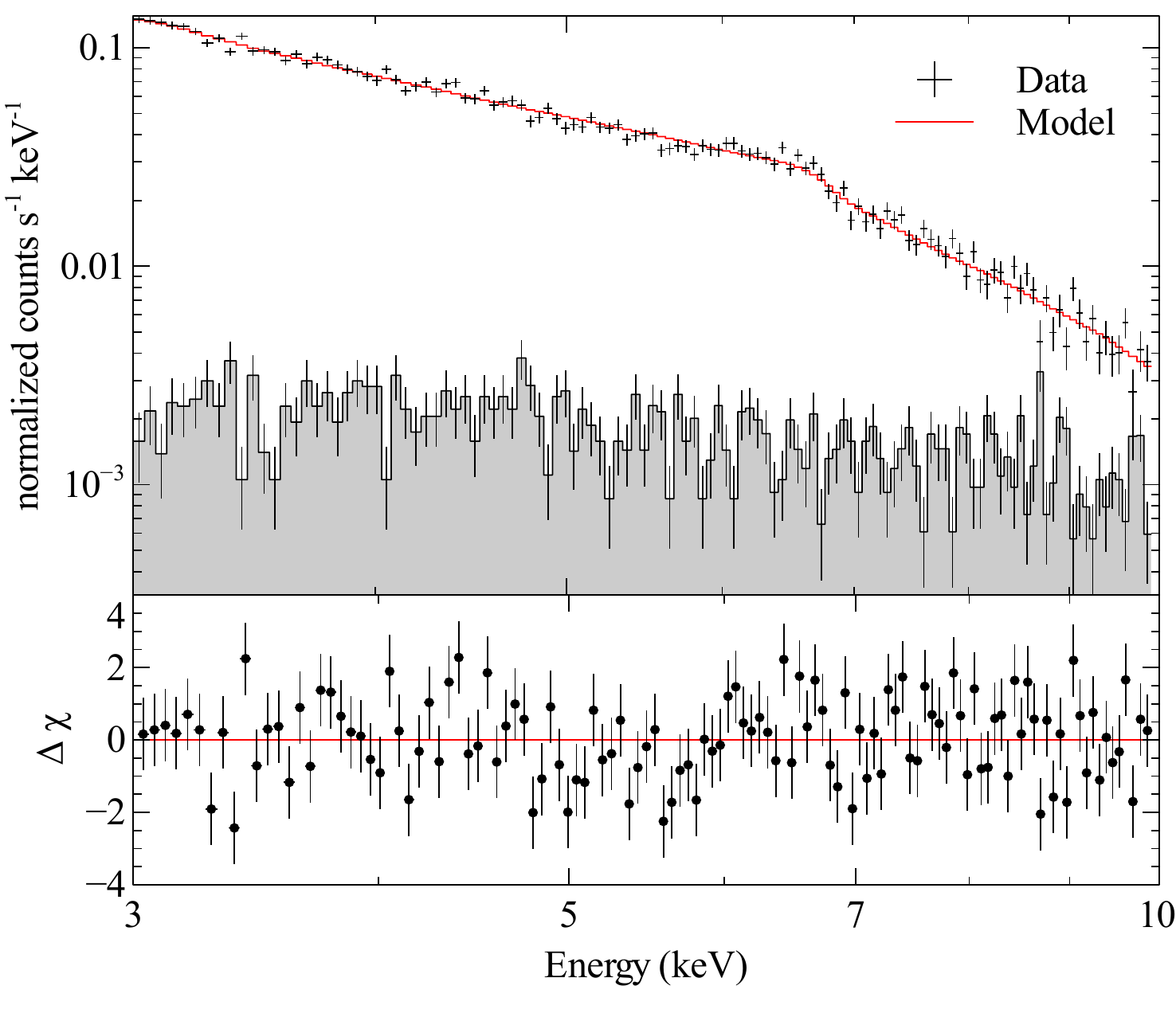}
\caption{Example fit to the 3--10~keV energy band, using the \emph{relxill} model. The top panel shows the data, model, and background (shaded region), and the bottom panel the residuals in units of standard deviations.}
\end{figure}

\begin{figure*}
\includegraphics[width=0.3\linewidth]{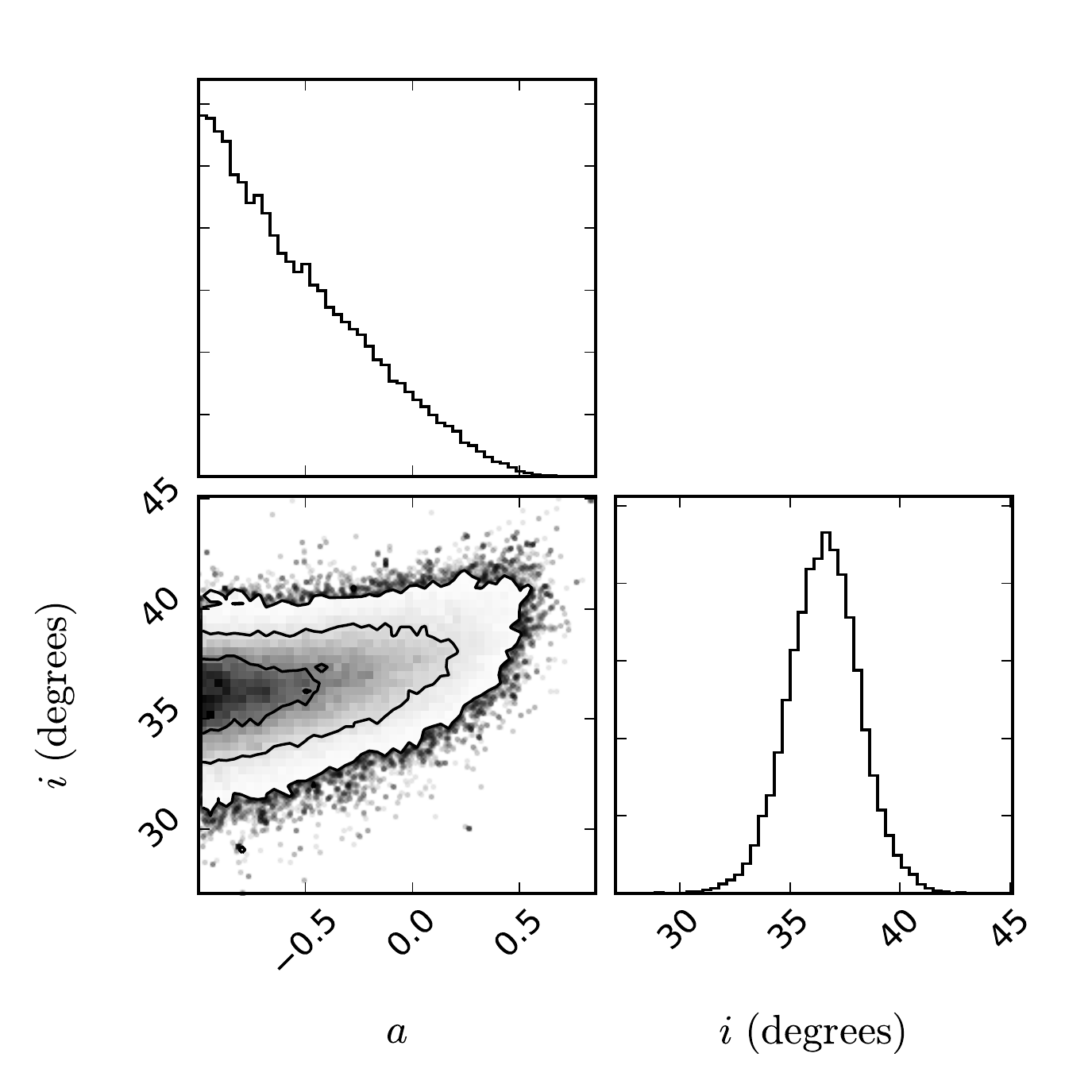}
\includegraphics[width=0.3\linewidth]{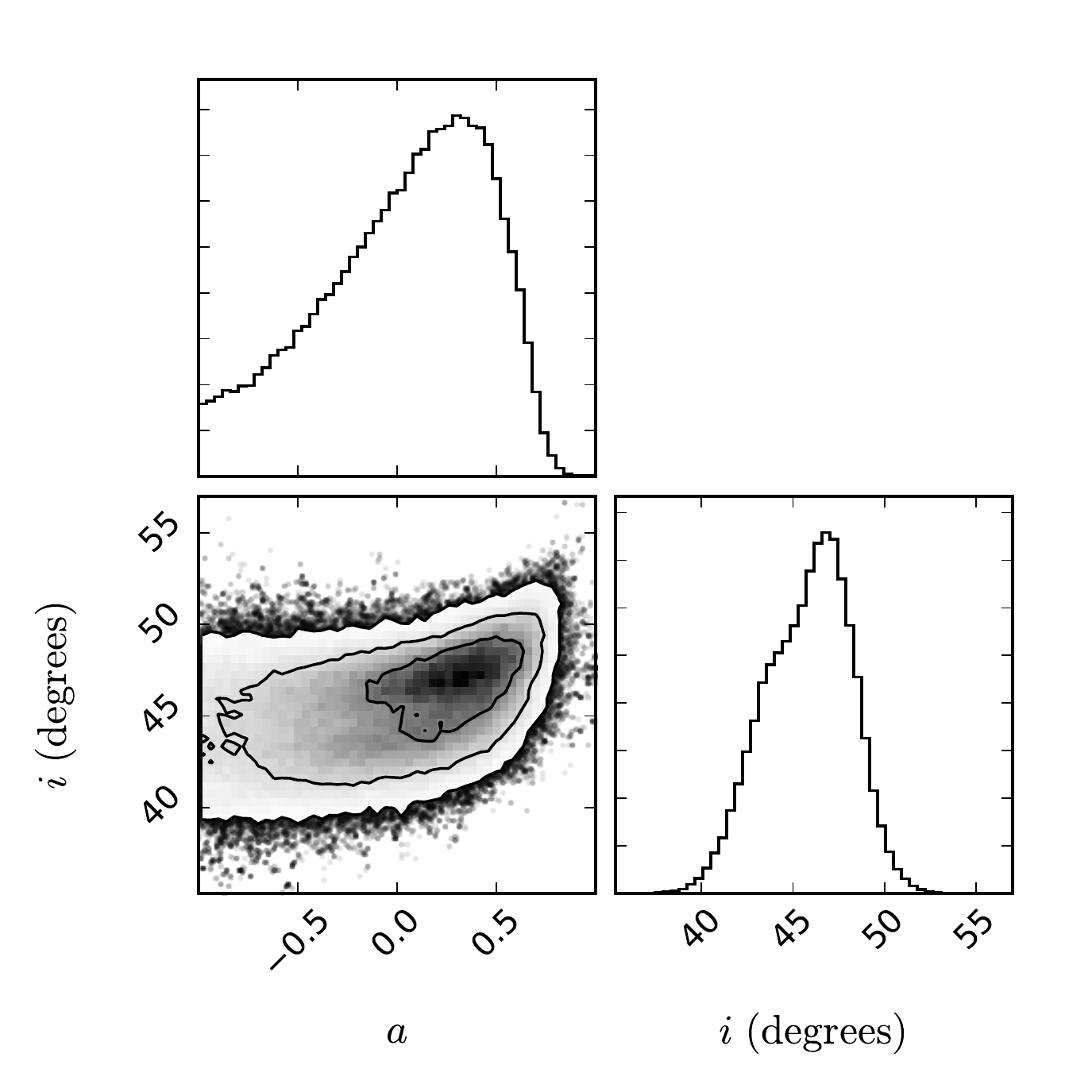}
\includegraphics[width=0.3\linewidth]{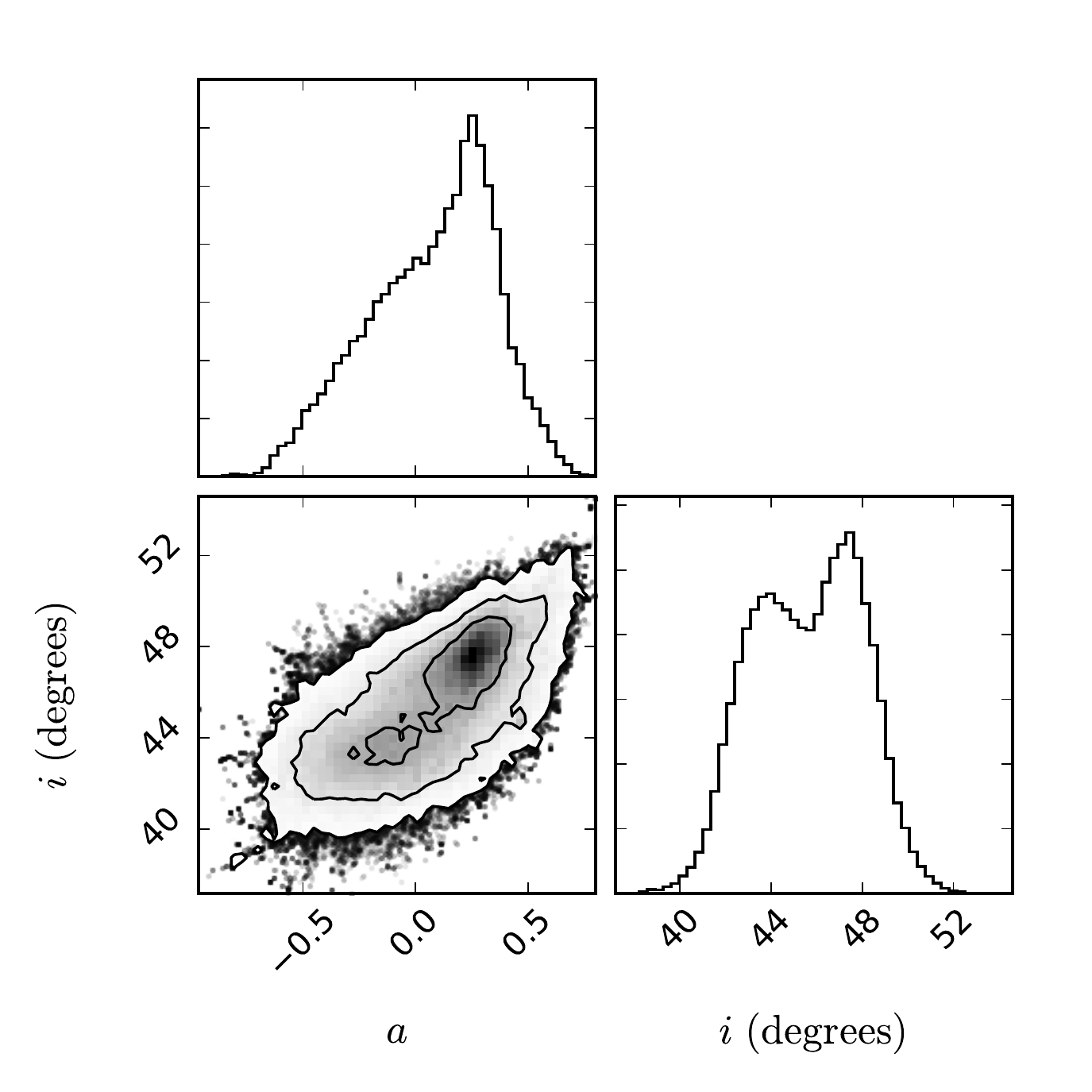}
\caption{MCMC contours for \emph{reflionx} (left), \emph{relxill} (middle) and \emph{relxilllp} (right). Contours are calculated from 10,000 points drawn randomly from the second half of the chains. In all cases, there is a mild degeneracy between spin and inclination, and maximal spin is ruled out at $>3\sigma$ confidence.}
\label{fig_mcmc}
\end{figure*}

When we fit the \emph{relxill} and \emph{reflionx} models with a free emissivity profile, they converge on an unphysical solution with retrograde spin and an extremely steep ($q=10$) emissivity index, due to the degeneracy between these parameters. An emissivity profile this steep is only possible with near maximal prograde spin, and even then only in the innermost regions of the disk \citep[e.g.][]{Wilkins12}. We therefore fix the index to the classical $q=3$ solution for these models. For \emph{relxilllp}, we keep the source height free, and tie the reflection fraction to that expected from the lamp-post geometry.

\begin{table*}
\begin{center}
\caption{Fit parameters for the fits to the iron line band. Errors are calculated from the marginalized parameter distributions from the MCMC analysis. We neglect the iron abundance, which is not meaningfully constrained, due to the strong degeneracy between reflection fraction and iron abundance in this band.}
\label{table_ironlinefits}
\begin{tabular}{l l c c c c c c c c c r}
\hline
\hline
Model & $a$ & $i$ & $\Gamma$ & $\log(\xi)$ & $h$ & $F_\mathrm{pl}$ & $F_\mathrm{ref}$ &$R$& $\chi^2/$dof & $n_\mathrm{steps}$\\
&&(degrees)&&(erg cm s$^{-1}$)&$R_\mathrm{G}$&\\
\hline
\textsc{reflionx}	&	$<-0.4$				&	$36\pm1$		&	$2.11\pm0.02$	&	$2.6\pm0.1$	&-& 2.79 & 0.21	&-&154/114	&	10000	\\
\textsc{relxill}	&	$0.3^{+0.1}_{-0.5}$	&	$47^{+1}_{-3}$	&	$2.28\pm^{+0.32}_{-0.06}$	&	$2.7_{-2.1}^{+0.1}$	& -	&-&3.00$^1$&	$2.2_{-0.4}^{+0.7}$&151/114	&	100000	\\
\textsc{relxilllp}	&	$0.3^{+0.1}_{-0.6}$	&	$44^{+3}_{-2}$	&	$2.30^{+0.23}_{-0.06}$	&	$2.68^{+0.03}_{-1.89}$ &	$<2.2$	&-&3.00$^1$&0.70$^2$& 146/113	&	200000	\\
\hline
\end{tabular}
\end{center}

\begin{flushleft}

Fluxes are in units of 10$^{-12}$~erg~cm$^{-2}$~s$^{-1}$, and are defined from 3--10~keV.\\
$^1$The power-law component flux in \emph{relxill} is defined by the reflection fraction. In these cases, the reflection flux $F_\mathrm{ref}$ refers to the total flux of the \emph{relxill} model. See \citet{Dauser16} for details of how the reflection fraction is implemented in \emph{relxill}.\\
$^2$The reflection fraction of \emph{relxilllp} is defined by the source height and disk inner radius, and therefore does not have independent errors.

\end{flushleft}
\end{table*}

In all three cases, we find similar results. Maximal spin is ruled out at $>3\sigma$ in all models, and the inclinations are around 40 degrees. We explore the uncertainty on spin using the \textsc{xspec\_emcee}\footnote{\url{github.com/jeremysanders/xspec\_emcee}} implementation of the \textsc{emcee} MCMC code \citep{Foreman-Mackey13}. For each of the three models, we use 100 walkers, burn the first 1000 steps and then run the chains until the Gelman-Rubin scale-reduction factor is $\leq1.2$ for all parameters. We calculate the errors on each parameter by marginalizing over the other parameters for the second half of each chain. In each case, there is a mild degeneracy between spin $a$ and inclination $i$, shown in Fig.~\ref{fig_mcmc}. However, even when this is taken into account, all three models rule out maximal spin at $>3\sigma$. The best fit model parameters are given in Table~\ref{table_ironlinefits}.

While fixing the emissivity profile to $q=3$ restricts it to a physical region of parameter space, it also risks biasing the spin. We therefore test manually changing the inner emissivity index for the \emph{relxill} model, with a break to $q=3$ at 6~$R_\mathrm{G}$, and find that generally the steeper the emissivity profile, the lower the spin, with maximal retrograde spin for $q=5$ or above. Alternatively, if we lower the emissivity to $q=2$, the best fit spin is maximal. This model is essentially similar to the high X-ray source model discussed below, where the emissivity index is relatively flat in the inner regions and breaks to $q=3$ further out.

Estimates of the inner radius of the accretion disk are known to be strongly degenerate with the height of the X-ray source \citep{Fabian14}, so it is possible that a high source height could mimic low spin. Our \emph{relxilllp} model has a very low source height and low spin, but the source height is most likely being set by the reflection fraction, which is fixed to the lamppost value. While fixing or constraining the reflection fraction can make the results more reliable in some cases \citep[e.g.][]{Dauser14,Parker14_mrk335}, in this case the reflection fraction is unreliable, as we are fitting a narrow energy band with no reference feature. To get around this, we instead fit with with the \emph{relline\_lp} model, which produces only a single line instead of the full reflection spectrum, and the source height is free to vary without being influenced by other parameters. In this case, we find that the spin is completely unconstrained, while the source height increases to $>250R_\mathrm{G}$, and the fit is equivalent in quality to that of the \emph{relxilllp} model. We therefore conclude that the low spin solution is not unique, and is degenerate with other models where the X-ray corona is far from the disk or the disk itself is truncated around a high spin black hole.

We also consider the possibility that the sharp edge to the emission line just below 7~keV is caused by unresolved absorption from highly ionized gas, producing Fe~\textsc{xxvi} absorption lines just below 7~keV. This would act to narrow and sharpen the line, artificially lowering the estimated spin. We model this using a Gaussian line, with the width fixed at $0.1$~keV, applied to the \emph{relxill} model. This improves the fit by $\Delta\chi^2=3$, for 2 additional degrees of freedom. The spin value does not change significantly, but the errors increase in size ($a=0.4^{+0.3}_{-0.7}$).

Finally, we consider the possibility that the iron line is produced by ionized, but not relativistically blurred, absorption. We fit the same data and energy band with the \emph{xillver} model \citep{Garcia10}, using the same procedure as the three relativistic models. We find a significantly worse fit: $\chi^2/$dof $=213/115=1.85$, and conclude that relativistic reflection is required.

\subsection{Broad-band}

\begin{figure*}
\centering
\includegraphics[width=0.7\linewidth]{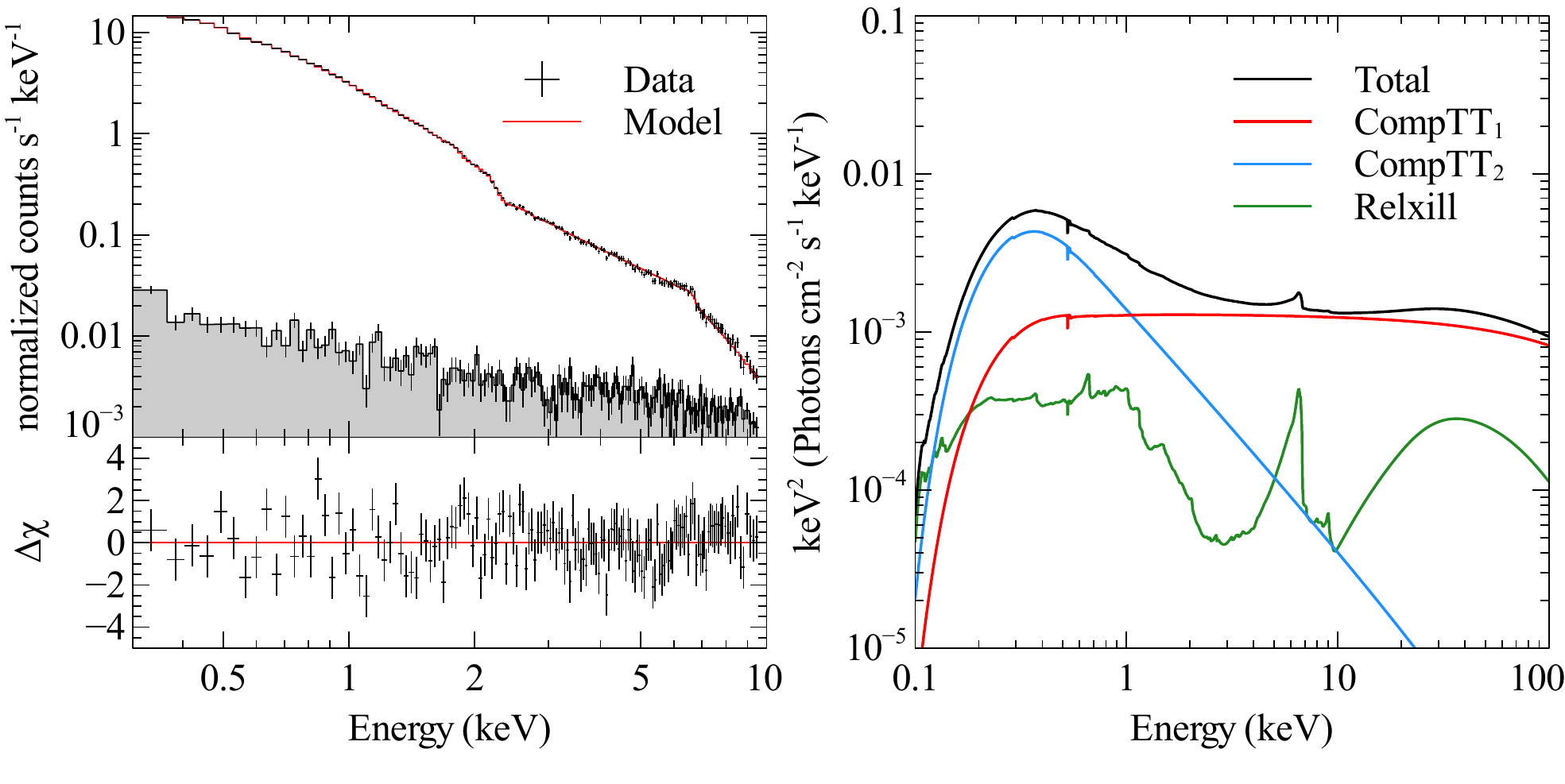}
\caption{Left: Data and residuals (in terms of standard deviations) for our best-fit broad-band model. Right: The different model components. Note the broad peak below 1~keV, which corresponds to the excess found in the data.}
\label{fig_broadband}
\end{figure*}

We use a phenomenological fit, similar to that in \citet{Fabian09}, for our first model of the broad-band spectrum. To avoid potentially biasing the fit by modelling the soft excess with reflection when it is due to another process, we use the Comptonization model \emph{CompTT} for both the continuum and soft excess, and fit the Fe~K$\alpha$ line using the relativistic emission line model \emph{relline}. This fit gives a reasonable description of the data, but leaves an emission feature around 1~keV. Because of the weakness of the feature and the low spectral resolution of the EPIC-pn at low energies it is hard to be certain of the origin of this feature. However, no narrow emission or absorption features are present at this energy in the RGS data, so it is likely to be broad.

We can model this feature using a broad emission line, with the blurring parameters tied to those of the Fe~K$\alpha$ line. In this case, the rest frame energy of the feature is 0.94$\pm0.02$~keV, with an equivalent width of 0.016~keV. The blurring parameters are consistent with those found from fitting the iron line alone ($a<0.7$, $i=47_{-2}^{+1}$ degrees), although this is likely because the low energy feature has minimal impact on the measured parameters. 

We next fit with a full reflection model, \emph{relxill}, replacing the two \emph{relline} components but retaining the two component Comptonization continuum. Because it is not clear which of the two continuum components (or what ratio of the two) is responsible for producing the reflected emission, we leave $\Gamma$ free in the reflection model within hard limits of 1.5 and 3, corresponding to the hard and soft continuum components. This model gives a reasonable fit ($\chi^2_\nu=167/142=1.17$, although there is a lot of scatter in the data points below 1.5, which contributes a large amount of the $\chi^2$ due to the high signal of these energy bins.
The best-fit parameters for this model are given in Table~\ref{table_broadband}. As in the narrow band case, the spin is low ($<-0.4$). Interestingly, in terms of spin and inclination, this solution looks more like the \emph{reflionx} solution from \S~\ref{subsec_ironband} than the \emph{relxill} or \emph{relxilllp} solutions. The reflection $\Gamma$ value found is intermediate between the indexes of the two Comptonization components \citep[e.g.][]{Vaughan02}, suggesting the reflection may be produced by a combination of the two.

\begin{table}
\centering
\caption{Best fit broad-band model parameters}
\label{table_broadband}
\begin{tabular}{l c c l}
\hline
\hline
Component & Parameter & Value & Description\\
\hline
\emph{comptt$_1$} & $T_0$    & $0.06\pm0.01$ 				& Seed temperature (keV)\\
				  & $T_1$    & $150^*$ 			   			& Plasma temp. (keV)\\
				  & $\tau_1$ & $0.12^{+0.03}_{-0.06}$		& Optical depth\\
				  & $F_\mathrm{0.5-10}$&	6.03			& 10$^{-12}$~erg~cm$^{-2}$~s$^{-1}$\\
\emph{comptt$_2$} & $T_2$    & $75_{-23}^{+31}$				& Plasma temp. (keV)\\
				  & $\tau_2$ & $<9$							& Optical depth\\
				  & $F_\mathrm{0.5-10}$&	4.10			& 10$^{-12}$~erg~cm$^{-2}$~s$^{-1}$\\
\emph{relxill}	  & $a$		 & $<-0.4$						& Spin\\
				  & $i$		 & $39_\pm2$					& Inclination (degrees)\\
				  & $\Gamma$ & $2.1\pm0.1$					& Photon index\\
				  & $\log(\xi)$& $3.0$					& Ionization (erg cm s$^{-1}$)\\
				  & $A_\mathrm{Fe}$ & $>9$ 					& Iron abundance (Solar)\\
				  & $F_\mathrm{0.5-10}$&	0.93			& 10$^{-12}$~erg~cm$^{-2}$~s$^{-1}$\\
\hline
\hline		
\end{tabular}
$^*$We fix the plasma temperature of the hard \emph{comptt} component to 150~keV, as it is outside the detector band, but fit for the soft component temperature in case more curvature is required.
\end{table}

We also investigate a more conventional double-reflection model, where the two reflection components are identical apart from different ionizations, which has been successfully used in other sources with strong soft excesses \citep[e.g.][]{Fabian09}. We model this with two relxill components and a cut-off power-law continuum, freeing the inner emissivity index and fixing the break radius to 6~$r_\mathrm{G}$. This model gives a significantly worse fit ($\chi^2$/dof$=352/145=2.43$), and maximal spin $a>0.997$. 


\section{Discussion}
\label{section_discussion}

The spin of supermassive black holes is set by their growth history, so the expected distribution of spin parameters depends on the mechanism by which they grow \citep[e.g.][]{Berti08, Fanidakis11}. Continuous accretion should spin up the black hole, whereas chaotic accretion would result in low spin. Additionally, merging black holes conserve their angular momentum at the point of merger, resulting in a relatively high (but not maximal) spin. If the low-spin interpretation of the Ton~S180 X-ray spectrum is correct, this is evidence for a wide range of spin values in AGN, implying that multiple mechanisms may be driving the growth of AGN. 

We cannot robustly conclude that the low-spin conclusion is correct, due to the strong degeneracy between this solution and two other cases: large source height, and a truncated accretion disk. Spin measurements using reflection rely on the assumption that the disk terminates close to the ISCO, so if the disk instead truncates at a larger radius a lower spin value will be found. Similarly, if the X-ray source is far from the disk, or closer and outflowing, this preferentially illuminates the outer disk, mimicking the effect of truncation \citep{Fabian14}. There is no obvious method to distinguish these models, other than if at some future time a smaller inner radius is measured. This would rule out low spin, but still not distinguish between truncation or an elevated X-ray source. Because the relativistic distortion is mild, we do not require a compact corona in this case. As the source height in the lamp post model is consistent with hundreds of gravitational radii, the corona could be similarly large.

Models using the lamp-post geometry have been criticised in recent work for their reliance on an unphysical geometry \citep[][]{Niedzwiecki16}: the X-ray emitting region cannot be due to a point source. While this is a fair criticism of these models if they are taken too literally (i.e. interpreted as direct evidence of a specific geometry, rather than more generally as evidence for a compact or extended source), it is not clear that they are any less valid than broken power-law models as a phenomenological tool for fitting the emissivity profile. However, using the reflection fraction from the lamp-post geometry (as is an option in \emph{relxilllp}) will bias the result, as this has a major impact on the spectrum and constrains the emissivity profile without relying on measuring the shape of the iron line \citep[see][for discussion on using the reflection fraction as a constraint on spin]{Dauser14, Parker14_mrk335}. The lamp-post reflection fraction is only valid for a compact source, whereas the lamp-post emissivity profile may well be a reasonable approximation for a range of geometries (just as the broken power-law is).
Our results here show that the lamp-post emissivity profile used in \emph{relxilllp} returns results that are entirely consistent with those of the power-law \emph{relxill}, as do the results of \citet{Parker14_mrk335} in the high-spin case of Mrk~335. We conclude that lamp-post models are still a suitable technique for measuring spin in AGN and X-ray binaries, unless they can be demonstrated to produce a qualitatively different result.

Previous authors have advocated for a partial-covering absorption origin for the iron line feature in AGN \citep[see review by][]{Turner09}. While this scenario is generally quite degenerate with reflection models from a purely spectral perspective, it struggles to explain the timing properties of AGN, in particular the presence of the broad iron line found in short timescale reverberation lags \citep[e.g.][]{Zoghbi12}.
Microlensing results \citep[e.g.][]{Chartas17} have shown that, in lensed quasars, specific energy bands in the iron line profile can me amplified, corresponding to magnification of the inner accretion disk by individual microlensing events. This obviously requires that the different energy bands of the line profile be spatially separate, which is expected in the reflection scenario, but is not possible in the partial covering model where the line is instead produced by the Fe~K absorption edge, which does not have different energies at different positions in an absorbing cloud. 

Relativistically broadened Fe~K lines are also found in X-ray binaries (XRBS) with almost identical spectra \citep[e.g.][]{Walton12}, implying an equivalent physical process for their production. In this case, there is generally no equivalent of the broad line region to produce the eclipsing events seen in AGN, so the lines cannot be due to Compton-thick absorption. While there are plentiful absorption features in XRB spectra and partial covering does occur, these are generally due to highly ionized, relatively low density gas from either the stellar wind \citep[e.g.][and references therein]{Grinberg15} or a disk wind \citep[e.g.][]{Miller15}, not the high density, low ionization gas required to produce a broad Fe feature. A partial-covering model for the broad iron line in GRS~1915+105 was proposed by \citet{Mizumoto16}, but does not address these issues, the state dependence of the broad Fe line, or the strong narrow Fe K$\alpha$ line that should be produced by scattering in this scenario and which is not found in the data.

In the case of Ton~S180, based on the lack of any detectable absorption in the RGS data, it appears that the system is completely unobscured. We therefore consider it extremely unlikely that there is a `hidden' partial-covering Compton-thick absorber producing the iron feature.

Ton~S180 was included in the sample of \citet{Iso16}, who attempt to explain the broad Fe~K lines and RMS spectra of 20 Seyfert galaxies observed with \suzaku\ using a double partial-covering model. This model requires an absorber with a column density of $4.5\times10^{24}$~cm$^{-2}$ covering 70--80\% of the X-ray source and an Fe~K$\alpha$ line fixed at 6.16~keV (the reason for which is not explained). The absorption models used are not appropriate for the Compton-thick regime, so it is not clear that the results are valid, and the work includes several sources for which strong Fe~K reverberation signals have been detected, a phenomenon which makes no sense in this interpretation. We conclude that while this model can explain the observed spectra (above 2~keV) and some of the simpler variability results, it is not physically self-consistent, and leaves many observed properties of the sources concerned unexplained.

Multiple suggestions have been proposed for the origin of the soft excess over the years, including relativistic reflection \citep{Crummy06}, absorption \citep{Middleton07}, or Comptonisation \citep{Czerny87}, and we note that it is entirely possible that multiple processes contribute to the soft excess. In some sources, fits which model the soft excess with reflection give entirely consistent results to fits which only include the iron line \citep[see e.g.][]{Fabian13_iras,Parker17_iras13224}, which argues in favour of the reflection interpretation of the soft excess in these sources. However, in the case of Ton~S180, the results we find from fitting just the iron line with reflection are inconsistent with those found previously by fitting both the iron line and soft excess \citep{Nardini12, Walton13_suzaku}\footnote{We note that, given the quality of data available at the time, it would likely have been impossible for these authors to constrain the spin from just the iron line.}. This argues against the reflection interpretation for the soft excess in this case. We recommend that, when sufficient data quality is available, independent fits to the iron line band be used as a consistency check to broad-band fitting \citep[this also applies to X-ray binaries, see][]{Parker15_cygx1}.
A related point of interest is the wide range of different lag-energy spectra observed for those sources with Fe~K reverberation lags \citep[e.g.][]{Kara13, Kara16_sample} -- not all sources with strong iron emission lags show a corresponding lag from the soft excess, implying that multiple emission mechanisms can contribute to the soft excess. 

We note that the best fit value for the photon index of the reflection spectrum is intermediate between the two continuum components. This may be an indication that both contribute a similar amount to the reflected emission, and thus are located close together, potentially in a stratified corona. There are various pieces of evidence that suggest there are a range of temperatures within the X-ray emitting region, notably propagation lags \citep[e.g.][]{Arevalo06} where the low-frequency lag properties of X-ray binaries and AGN are explained by fluctuations propagating through an extended corona, causing a delay between different energy bands. It is not obvious why there would be such a large amount of spectral curvature in Ton~S180, but we can speculate that this is related to the low spin (or large truncation radius, or elevated corona). If the X-ray emitting region is correspondingly larger, it could have a larger range of temperatures, optical depths etc., leading to a more curved continuum. The obvious way to test this is with timing analysis, and we note that \citet{Arevalo14} found that the timing properties of NGC~3227 were consistent with two power-law/Comptonization components, one cold and slowly variable, the other hotter and more rapidly variable. The spectral modelling of this source by \citet{Markowitz09} found similar results to those we find here, although in that case the two-component continuum was strongly modified by multiple absorption zones.

Our best-fit broadband model strongly prefers super-solar iron abundance ($A_\mathrm{Fe}>9$). This is caused by the strong iron line and relative absence of other reflection features. One possibility for mitigating this is that the reflection spectrum should actually be harder than in our best fit model, as the iron abundance is somewhat degenerate with the photon index. Because we have no hard-band coverage, and so cannot measure the Compton hump, it is difficult to be sure of the exact shape of the reflection spectrum. Super-solar abundances have been estimated with reflection modelling in many AGN previously, \citep[e.g.][]{Fabian09, Walton13_suzaku, Risaliti13, Parker14_mrk335}, and it is possible that the apparent iron abundance of the inner accretion disk can be enhanced by various processes, notably radiative levitation \citep{Reynolds12}. More generally, the abundance found here is not necessarily representative of the iron abundance in the mid-plane of the disk, the overall accretion flow, or the host galaxy. Alternatively, our two-component continuum may be affecting the measured parameters. It may be that this is only a phenomenological model for a more complex curved continuum, leading to the reflection model being adjusted to compensate.

Finally, differences in the density of the disk can potentially explain the extremely high iron abundances \citep{Garcia16}. In a forthcoming paper (Jiang et al., in prep) we will show that these models can drastically lower the observed iron abundance.
We note that \citep{Takahashi10} find iron abundances around solar in their fit, which uses the \emph{reflionx} model. The reason for this difference is not obvious, but it appears that the \xmm\ data prefers a smaller reflection contribution to the soft excess, which drives the iron abundance higher to produce the line.

\section{Conclusions}
\label{section_conclusions}
We have fit the \xmm\ EPIC-pn and RGS spectra of the longest archival observation of the rapidly variable NLS1 Ton~S180, finding several interesting results:
\begin{itemize}
\item We fit the iron band with the \emph{reflionx}, \emph{relxill}, and \emph{relxilllp} relativistic reflection models. With all three models we find low spin, with maximal spin ruled out at $>3\sigma$. This interpretation is then confirmed by our broad-band fit, using a two-component Comptonization continuum to produce the observed spectral curvature.
\item We find good agreement between between \emph{relxill} and the lamp-post geometry variant \emph{relxilllp}, suggesting that using this geometry does not bias the results of spin measurements.
\item We cannot exclude the possibility that the low spin we estimate is instead due to a large source height, an outflowing corona, or truncation of the accretion disk before the ISCO. While it could be easily demonstrated that the source has high spin by taking another measurement where a much broader line is found, it is much harder to conclusively prove the low spin value.
\end{itemize}

\section*{Acknowledgements}
MLP is supported by a European Space Agency (ESA) Research Fellowship.
MLP and ACF acknowledge support from the European Research Council through Advanced Grant on Feedback 340492. 

\bibliographystyle{mn2e}

\end{document}